\documentstyle[preprint,aps]{revtex}
\begin{document}
\draft
\title{Thermal properties of $\bbox{{}^{54}}$Fe}

\author{D. J. Dean${}^1$, S. E. Koonin${}^1$, K. Langanke${}^1$,
P. B. Radha${}^1$, and Y. Alhassid${}^2$}
\address{${}^1$W. K. Kellogg Radiation Laboratory, 106-38, California
Institute of Technology\\ Pasadena, California 91125 USA\\
${}^2$Center for Theoretical Physics, Sloane Physics Laboratory\\
Yale University, New Haven, Connecticut, 06511 USA}
\date{\today}
\maketitle

\begin{abstract}
We study the thermal properties of
${}^{54}$Fe with the Brown-Richter interaction in the complete $1p0f$
model space.
Monte Carlo calculations show a peak in the heat capacity and
rapid increases in both the moment of inertia and
$M1$ strength
near a temperature of 1.1~MeV that are associated with the
vanishing of proton-proton and neutron-neutron monopole pair
correlations; neutron-proton correlations persist to higher
temperatures. Our
results are consistent with a Fermi gas level density whose
back-shift
vanishes with increasing temperature.
\end{abstract}

\narrowtext

The nuclear level density increases rapidly at excitation energies
above several MeV and it becomes difficult to resolve or calculate
individual states. In this regime, it is more appropriate to employ a
statistical description where observables are averaged over the many
levels at a given energy. The concept of an equilibrated compound
nucleus is among the most fundamental of nuclear reaction theories
\cite{Bohr}, and plays a central role in our understanding of
processes induced by probes ranging from photons to heavy ions.

While the proper description of a compound nucleus is in terms of a
microcanonical (fixed-energy) ensemble, it is often more convenient
to consider a canonical ensemble whose temperature is chosen to
reproduce the average excitation energy. In the past decade, there
has been renewed experimental \cite{exp} and theoretical
\cite{theory} effort to explore the properties of heavy nuclei at
finite temperature and high spin. The properties of hot nuclei are
also important in various astrophysical scenarios, particularly in
the late stage of a supernova collapse and explosion \cite{aufder}.

Most theoretical approaches to hot nuclei devolve to a mean-field
description based on an average configuration \cite{goodman}. The
realization that thermal and quantal fluctuations about the average
can be important has prompted more sophisticated approximations
\cite{alhassid}, although even these have clear limitations. In
principle, the nuclear shell model (which provides a complete
spectrum and wavefunctions) offers a fully microscopic approach to
the problem. However, conventional finite-temperature shell model
calculations within a
complete major shell are limited to light nuclei (${}^{20}$Ne and
${}^{24}$Mg) in the $sd$ shell \cite{Miller}.

In this Letter, we exploit recently developed Monte Carlo techniques
to calculate the thermal properties of ${}^{54}$Fe in a complete
$0\hbar\omega$ model space with a realistic interaction. The methods
we use describe the nucleus by a canonical ensemble at temperature
$T=\beta^{-1}$ and employ a Hubbard-Stratonovich linearization of the
imaginary-time many-body propagator, $e^{-\beta H}$, to express
observables as path integrals of one-body propagators in fluctuating
auxiliary fields \cite{mcrefb}. Since Monte Carlo techniques
avoid an explicit enumeration of the many-body states, they can be
used in model spaces far larger than those accessible to conventional
methods. The Monte Carlo results are in principle exact and are in
practice subject only to controllable sampling and discretization
errors. The nucleus we have chosen for this initial study
(${}^{54}$Fe) is among the most abundant in the presupernova core of
a massive star, so that its thermal properties are of considerable
astrophysical import. Further, the ground states of nuclei in the
mid-$pf$-shell are dominated by nucleon-nucleon correlations (e.g.,
pairing) whose evolution with increasing temperature is of particular
interest.

To circumvent the ``sign problem'' encountered in the Monte
Carlo shell model calculations with realistic interactions,
Alhassid {\it et al.}
\cite{mcrefa} suggested an extrapolation procedure from a family of
Hamiltonians that are free of the sign problem
to the physical Hamiltonian. One defines a set of Hamiltonians
$H_g=H_G+gH_B$ such that $H_{g=1}=H$ is the physical Hamiltonian and
$H_{G,B}$ are the ``good'' and ``bad'' parts of the Hamiltonian,
respectively. For $g\leq0$, $H_g$ is free of the sign problem and
calculated observables are extrapolated to $g=1$. For ground state
properties, this procedure was validated by comparison to direct
diagonalization results in the $sd$ and lower $pf$ shells. However,
it is impractical at intermediate temperatures due to the overly
strong pairing interaction in $H_g$ for $g<0$, which suppresses the
population of excited states. This problem can be corrected by
scaling $H_G$ as $\left(1-{1-g\over\chi}\right)H_G$, together with a
$g$-dependent compression of the single-particle spectrum; the value
of $\chi$ is chosen to make the $g$-extrapolation as smooth as
possible. Note that, as before, the original Hamiltonian is recovered
for $g=1$.

Our calculations include the complete set of
$1p_{3/2,1/2}0f_{7/2,5/2}$ states interacting through the realistic
Brown-Richter Hamiltonian \cite{rvjb}. Some $5\times10^9$
configurations of the 8 valence neutrons and 6 valence protons moving
in these 20 orbitals are involved in the canonical ensemble.
The results presented below have been obtained in MC shell model
studies with a time step of $\Delta\beta=1/32~{\rm
MeV}^{-1}$ using 5000--9000 independent Monte Carlo samples at seven
values
of the coupling constant $g$ spaced between $-1$ and 0 and the value
$\chi=4$.
A linear
extrapolation to the physical case {($g=1$)} is justified by
the quality-of-fit for most of the observables discussed below,
although
the quadrupole moments warranted a quadratic extrapolation.
We have tested our procedure for the $fp$-shell nucleus ${}^{44}$Ti
against a calculation performed with
the direct diagonalization code CRUNCHER \cite{Ressell} and were able
to reproduce the excitation energy
as a function of temperature for the temperature interval relevant
for
this paper \cite{tobe}.
However, we note that an exact reproduction of the CRUNCHER
energies is obtained only after a $\Delta \beta \rightarrow 0$
extrapolation, which
lowers the absolute energies slightly compared to the calculation
with the finite
value $\Delta\beta=1/32~{\rm MeV}^{-1}$. For ${}^{54}$Fe we have
checked
at several temperatures
that our qualitative results are not
changed by the $\Delta\beta$ extrapolation. We have also checked that
using $\chi=3$ in the $g$-extrapolation
does not change our results.

The calculated temperature dependence of various observables is shown
in Fig.~\ref{fig1}.
In accord with general thermodynamic
principles, the internal energy $U$ steadily increases
with increasing temperature \cite{foot}. It shows an inflection point
around
$T \approx 1.1$~MeV, leading to a peak in the heat capacity, $C\equiv
dU/dT$, whose physical origin we will discuss below.
The decrease in $C$ for $T \gtrsim 1.4$~MeV
is due to our finite model space (the Schottky effect
\cite{schot}); we estimate that limitation of the model space to only
the $pf$-shell renders our calculations of ${}^{54}$Fe quantitatively
unreliable for
temperatures above this value (internal energies $U\gtrsim 15$~MeV).
The same behavior is apparent in the level density parameter,
$a\equiv C/2T$.
The empirical value for $a$ is
$A/8~{\rm MeV} =6.8~{\rm MeV}^{-1}$ which is in good agreement with
our results for
$T \approx 1.1$--1.5~MeV.

We also show in Fig.~\ref{fig1} the expectation values
of the squares of the $J=0$ proton-proton and neutron-neutron
pairing fields,
$\langle\Delta^\dagger\Delta\rangle$. Although the pair
wave function we have used (the BCS form, in which all time-reversed
pairs have equal amplitudes) is not necessarily optimal, these
observables are a rough measure of the number of $J=0$ pairs in the
nucleus. At low temperatures, the pairing fields are significantly
larger than those calculated for a non-interacting Fermi gas,
indicating a strong coherence in the ground state. With increasing
temperature, the pairing fields decrease;
both approach the Fermi gas values for $T\approx 1.5$~MeV and follow
it closely for even higher temperatures. Associated with the breaking
of pairs is a dramatic increase
in the moment of inertia, $I$, for $T=1.0$--1.5~MeV; this is
analogous to the
rapid increase in magnetic susceptibility in a superconductor. At
temperatures above 1.5~MeV, $I$ is in agreement with the rigid
rotor value, $10.7\hbar^2$/MeV; at even higher temperatures it
decreases
linearly due to our finite model space.

In Fig.~\ref{fig2}, we show various static observables. The $M1$
strength
unquenches rapidly with heating near the transition temperature.
However, for $T=1.3$--2~MeV $B(M1)$
remains significantly lower than
the single-particle estimate ($41~\mu_N^2$),
suggesting a persistent quenching at temperatures above the
like-nucleon depairing. This finding is supported by
the near-constancy of the Gamow-Teller $\beta^+$ strength, $B({\rm
GT}_+)$, for temperatures up
to 2~MeV. As the results of Ref.~\cite{Dean} demonstrate that
neutron-proton correlations are
responsible for much of the GT quenching in iron nuclei at zero
temperature, we interpret the present
results as evidence that isovector proton-neutron correlations
persist to higher temperatures. We have verified
that, in our restricted model space, both the ${\rm GT}_+$ and $M1$
strengths
unquench at temperatures above 2~MeV and that, in the
high-temperature limit, they both
approach the appropriate Fermi gas values. We note
that it is often
assumed in astrophysical calculations that the GT strength is
independent of temperature \cite{aufder}; our calculations
demonstrate that this is true for the relevant temperature regime
($T<2$~MeV).
We also note that a detailed examination of
the occupation numbers of the various orbitals
show no unusual
variation as the pairing vanishes.

The isoscalar mass quadupole moment,
$\langle Q^2\rangle$, increases at temperatures near the phase
transition (Fig.~\ref{fig2}, upper right), while
the isovector moment,
$\langle Q_{\rm v}^2\rangle$, decreases as the nucleus is heated,
showing a
minimum near $T=1.1$~MeV (Fig.~\ref{fig2}, lower right). The
behaviors of
$\langle Q^2\rangle$ and $\langle Q_{\rm v}^2\rangle$ imply that
$\langle Q_p \cdot Q_n \rangle= (\langle Q^2\rangle- \langle Q^2_{\rm
v}\rangle)/4$
increases near the phase transition
($Q_{p,n}$ are
the proton and neutron quadrupole moments).
Noting the relation
between
$\langle Q_p \cdot Q_n \rangle$ and the orbital part of the $M1$
strength
\cite{Richter}, we interpret the partial unquenching of the $B(M1)$
near $T=1.1$~MeV as related to the orbital part, while the spin part,
dominated by the same operator as the GT strength, remains
significantly
quenched to higher temperatures.

We have compared our results for $U$
with two simple models. As in
Ref.~\cite{Davidson}, we define the partition function of the nucleus
as
\begin{equation}
Z=\sum_i(2J_i+1) e^{-\beta E_i}+ \int^\infty_{E_0} \rho(E) e^{-\beta
E} dE\;,
\end{equation}
where the sum runs over the experimentally known nuclear levels and
the continuum state density $\rho(E)$ has been approximated by the
backshifted
Fermi gas model \cite{Thielemann}, where the backshift $P$ accounts
for the energy to break a pair; we
adopted a level density parameter $a=7.2~{\rm MeV}^{-1}$ and chose
$E_0=4$~MeV to smoothly match the two terms in Eq.~(1).
The first model had the conventional $T$-independent
backshift $P_0=1.45$~MeV \cite{Thielemann}, while the second model
simulated the vanishing of the pairing by a temperature-dependent
$P$:
\begin{equation}
P(T)=P_0 \left(1+ \exp\left\{ T-T_0\over\alpha\right\}
\right)^{-1}\;.
\label{neweqn}
\end{equation}
Our choice of the parameters $T_0= 1.05$~MeV, $\alpha=0.25$~MeV was
motivated from the two upper right panels in Fig.~\ref{fig1}. As seen
from the solid curves in the left panels of Fig.~\ref{fig1}, the
$P(T)$ model is in better agreement with
the Monte Carlo shell model results than is the constant-$P$ model;
in particular, there are clear maxima in both the heat capacity and
the level density
parameter related to the pairing phase transition. The dashed curves
in these panels indicate that the assumption of a
constant level density parameter is not quite appropriate at
temperatures below the phase transition.

In conclusion, we have demonstrated that shell model Monte Carlo
methods are well-suited to studying the finite-temperature properties
of nuclei using realistic effective two-body interactions.
Our calculations of ${}^{54}$Fe in the complete $pf$-shell show clear
signatures of a pairing phase transformation at a temperature of
1.1~MeV,
but persistent quenching of the Gamow-Teller $\beta^+$ strength at
higher temperatures. Results at temperatures above 1.5~MeV will
become reliable only when two or more major shells are included in
the calculations, an elaboration that is computationally quite
feasible. The extension of these calculations to other
interactions, heavier nuclei, and other observables should allow a
more thorough understanding of nuclear properties at high excitation
energies than is now possible by other methods.

\acknowledgements
This work was supported in part by the National Science Foundation,
Grants No. PHY90-13248 and PHY91-15574, and the Department of Energy,
Contract No. DE-FG-0291-ER-40608. We are grateful to P.~Vogel and
W.~E.~Ormand for
helpful discussions and to an anonymous referee for constructive
criticism. Computational cycles were provided by the Concurrent
Supercomputing Consortium and by the VPP500, a vector parallel
processor at the RIKEN supercomputing facility; we thank
Drs.~I.~Tanihata and S.~Ohta for their assistance in using the
latter.

\begin{figure}
\caption{Temperature dependence of various observables in
${}^{54}$Fe. Monte Carlo points with statistical errors are
shown at each temperature $T$. In the left-hand
column, the internal energy, $U$, is calculated as $\langle H \rangle
-E_0$, where $H$ is the many-body Hamiltonian and $E_0$ the ground
state energy. The heat capacity $C$ is calculated by a
finite-difference approximation to $dU/dT$, after
$U(T)$ has been subjected to a three-point smoothing, and the level
density parameter is $a\equiv C/2T$. The dashed and solid curves in
these panels correspond to the constant- and
temperature-dependent-backshift Fermi gas models, as described in the
text. To eliminate the systematic error associated with the
determination of $E_0$, we have chosen this parameter so that the
Monte Carlo and Fermi gas results for $U$ are equal at $T=0.66$~MeV.
In the right-hand column, we show the expectation values of the
squares of the proton and neutron pairing fields, where
 $\Delta^+_p= \sum p^\dagger_{jm}
p^\dagger_{j\bar m}$ (and similarly for neutrons) and the sum is
over all orbitals with $m>0$. For comparison, the
pairing fields calculated in an uncorrelated Fermi gas are
shown by the solid curve. The moment of
inertia is obtained from the expectation values of the square of the
total angular momentum by
$I=\beta \langle J^2 \rangle/3$.}
\label{fig1}
\end{figure}

\begin{figure}
\caption{The upper left panel shows the total magnetic dipole
strength, $B(M1)$ in units of nuclear magnetons; it is calculated
using free-nucleon
$g$-factors. The lower left panel shows the GT $\beta_+$ strength.
The low-temperature value of $4.7\pm0.2$ is somewhat larger than the
$4.3\pm0.2$ given in Refs.~[9,15], since the latter calculations were
done with $\Delta\beta=1/16~{\rm MeV}^{-1}$. Calculations at
$\Delta\beta=1/64~{\rm MeV}^{-1}$ indicate that the present
low-temperature value is converged. The upper and lower right panels
show the isoscalar ($Q= Q_p + Q_n$) and isovector ($Q_{\rm v} =Q_p -
Q_n$) quadrupole strengths, where  the quadrupole operators are
$r^2Y_2$
and results are given in terms of the oscillator
length, $b=1.96$~fm.
}
\label{fig2}
\end{figure}


\begin{references}
\bibitem{foot} After $\Delta\beta$ extrapolation, we calculate
$\langle H\rangle= -56.6\pm0.3$~MeV at $T=0.5$~MeV. Assuming that
$U(T)=aT^2$ at low temperatures and approximating $a$ by $a=3~{\rm
MeV}^{-1}$ (see Fig.~\protect\ref{fig1}), our calculation corresponds
to a ground state energy of $-57.3\pm0.3$~MeV.  Note that the ground
state energy quoted in Ref.~\protect\cite{mcrefa} had not been
extrapolated to $\Delta\beta=0$ and $T=0$.
\end{references}
\end{document}